\documentclass[amstex,twocolumn,showpacs,floats,floatfix,superscriptaddress,aps,pra]{revtex4}
\usepackage{amssymb}
\usepackage{amsmath}
\usepackage{calc}
\usepackage{graphicx}
\usepackage{bm}

\def\be{ \begin{equation} }
\def\ee{ \end{equation} }
\def\bea{ \begin{eqnarray} }
\def\eea{ \end{eqnarray} }
\def\bse{ \begin{subequations} }
\def\ese{ \end{subequations} }

\input{tcilatex}

\begin{document}

\author{G. S. Vasilev}
\affiliation{Department of Physics, Sofia University, James Bourchier 5 blvd, 1164 Sofia,
Bulgaria}
\author{P. A. Ivanov}
\affiliation{Department of Physics, Sofia University, James Bourchier 5 blvd, 1164 Sofia,
Bulgaria}
\author{N. V. Vitanov}
\affiliation{Department of Physics, Sofia University, James Bourchier 5 blvd, 1164 Sofia,
Bulgaria}
\affiliation{Institute of Solid State Physics, Bulgarian Academy of Sciences,
Tsarigradsko chauss\'{e}e 72, 1784 Sofia, Bulgaria}
\title{Exact solution of the optical Bloch equation for the Demkov model}
\date{\today }

\begin{abstract}
An exact analytic solution is presented for coherent resonant excitation of
a two-state quantum system driven by a time-dependent pulsed external field
described by Demkov model in the presence of dephasing.
\end{abstract}

\maketitle


\section{Introduction}


Coherent excitation influenced by dephasing processes represents an
important topic in quantum mechanics \cite{B.Shore},\cite{A-E}. Applications
of such models are numerious ranging from coherent atomic excitation and
quantum information to chemical physics and solid-state physics. Although a
significant effort have been devoted for studding the Bloch equations
corresponding to specific two-state models, almost all results are related
to some asymptotic regimes as weak dephasing, strong coupling or other
limits \cite{Vitanov}. There are very few exact solutions for the Bloch
equation. The complexity of this problem is due to the difficulty of
deriving an exact solution for third order linear differential equations. In
the case of resonant coherent excitation of a two-state system in the
presence of dephasing, solution can be found in \cite{Kyoseva}.

The original Demkov have been introduced in the theory of atomic collisions
\cite{Demkov}.


\section{Demkov model in the presence of dephasing\label{Sec-background}}

Dephasing processes can be incorporated into the description of resonant
excitation by including a phenomenological dephasing rate $\Gamma =1/T_{2}$,
where $T_{2}$ is the transverse relaxation time, into the Bloch equation,
\begin{equation}
\frac{\text{d}}{\text{d}t}\left[
\begin{array}{c}
u(t) \\
v(t) \\
w(t)%
\end{array}%
\right] =\left[
\begin{array}{ccc}
-\Gamma & -\Delta & 0 \\
\Delta & -\Gamma & -\Omega (t) \\
0 & \Omega (t) & 0%
\end{array}%
\right] \left[
\begin{array}{c}
u(t) \\
v(t) \\
w(t)%
\end{array}%
\right] ,  \label{Bloch-eq}
\end{equation}%
where the components of the Bloch vector $\left[ u(t),v(t),w(t)\right] ^{T}$
are expressed via density matrix elements $\rho _{mn}$ $\left(
m,n=1,2\right) $, as follows

\begin{eqnarray}
u(t) &=&2\text{Re}\rho _{12}(t)  \label{Bloch-vector} \\
v(t) &=&2\text{Im}\rho _{12}(t)  \notag \\
w(t) &=&\rho _{22}(t)-\rho _{11}(t)  \notag
\end{eqnarray}%
Hereafter the language of laser-atom interactions will be used, although the
results apply to any two-state system.The detuning $\Delta =\omega
_{0}-\omega $ is the difference between the transition frequency $\omega
_{0} $ and the carrier laser frequency $\omega $. The time-varying Rabi
frequency $\Omega (t)=\left\vert dE(t)\right\vert /\hbar $ describes the
laser-atom interaction, where $d$ is the electric dipole moment for the $%
\psi _{1}\leftrightarrow \psi _{2}$ transition and $E(t)$ is the laser
electric field envelope. For the Demkov model we have

\begin{eqnarray}
\Delta &=&\text{const,}  \label{Demkov-model} \\
\Omega (t) &=&\Omega _{0}\exp (-\left\vert t\right\vert /T),  \notag
\end{eqnarray}

\begin{equation*}
\Gamma =\text{const}
\end{equation*}%
The constant dephasing rate $\Gamma $ is a positive constant, and $T$ is the
characteristic pulse width. The peak Rabi frequency $\Omega _{0}$ will be
assumed also positive without loss of generality. For $\Gamma =0$, the Bloch
equation Eq.(\ref{Bloch-eq}) is solved exactly and this solution represents
the famous Demkov model \cite{Demkov} introduced in the theory of atomic
collisions.

We shall solve Eq.(\ref{Bloch-eq}) with the initial conditions corresponding
to a system initially in state $\left\vert 1\right\rangle $ i.e. $\rho
_{11}(-\infty )=1$ and $\rho _{22}(-\infty )=0.$ This corresponds to%
\begin{equation}
u(-\infty )=v(-\infty )=0,\text{ \ \ }w(-\infty )=-1.  \label{u-v-w-initial}
\end{equation}%
Our objective is to find the Bloch vector $\left[ u(t),v(t),w(t)\right] ^{T}$
and particulary, the population inversion $w(+\infty )$


\section{Analytic solution of the Demkov model}


Due to the specific form of the Demkov model, it is necessary to consider
the following two cases: $t\in I_{1}(-\infty ;0]$ and $t\in I_{2}[0;+\infty
).$ Let us begin with the first of them. Using Eq.(\ref{Demkov-model}) from
the Bloch system, Eq.(\ref{Bloch-eq}) we obtain third order differential
equation for the population inversion $w,$ which reads.%
\begin{gather}
\dddot{w}_{1}-2(T^{-1}-\Gamma )\ddot{w}_{1}+\left[ (T^{-1}-\Gamma
)^{2}+\Delta ^{2}+\Omega _{0}^{2}e^{2t/T}\right] \dot{w}_{1}+  \label{w-eq}
\\
\Omega _{0}^{2}(T^{-1}+\Gamma )e^{2t/T}w_{1}=0  \notag
\end{gather}%
A subscript "$1$" in the notation for the population inversion $w_{1}$
indicates that the Eq.(\ref{w-eq}) above and all formulas hereafter concerns
the time interval $t\in I_{1}(-\infty ;0].$ The solution of Eq.(\ref{w-eq})
can be expressed in terms of the generalized hypergeometric function $%
_{1}F_{2}(a_{1};b_{1},b_{2};x).$ Using the transformation

\begin{equation}
x=-\frac{1}{4}(T\ \Omega _{0})^{2}e^{2t/T}  \label{x-t}
\end{equation}

Eq. (\ref{w-eq}) is transformed to the following form%
\begin{gather}
x^{2}w_{1}^{^{\prime \prime \prime }}+x(2+T\ \Gamma )w_{1}^{^{\prime \prime
}}+  \label{w(x)-eq} \\
\left[ T\ \Gamma +\frac{(1-T\ \Gamma )^{2}+\Delta ^{2}}{4}-x\right]
w_{1}^{^{\prime }}-\frac{(1+T\ \Gamma )}{2}w_{1}=0  \notag
\end{gather}

Generalized hypergeometric function (GHF) $_{1}F_{2}(a_{1};b_{1},b_{2};x)$
satisfies the equation \cite{Rainville}%
\begin{equation}
x^{2}F^{^{\prime \prime \prime }}+(b_{1}+b_{2}+1)xF^{^{\prime \prime
}}+(b_{1}b_{2}-x)F^{^{\prime }}-a_{1}F=0,  \label{GHF-eq}
\end{equation}%
where $F$ is shortened notation for $_{1}F_{2}(a_{1};b_{1},b_{2};x)$. More
details regarding basic definitions and formulas for GHF are placed in sec.
Appendix. By comparing Eq.(\ref{w(x)-eq}) and Eq.(\ref{GHF-eq}), it is
trivial algebra to determine parameters $a_{1};b_{1},b_{2}$ of the GHF

\begin{equation}
b_{1}=\frac{1}{2}+\frac{T\ \Gamma }{2}+i\frac{T\ \Delta }{2},\
b_{2}=(b_{1})^{\ast },\ a_{1}=\text{Re}(b_{1}),  \label{a1-b1-b2}
\end{equation}%
where as usual the notation "$\ast $" stands for complex conjugation. By
reason to simplify the writing of the formulas, hereafter we use

\begin{equation}
\gamma =\frac{T\ \Gamma }{2};\ \delta =\frac{T\ \Delta }{2};\ \omega =\frac{%
T\ \Omega _{0}}{2}.  \label{g-d-o}
\end{equation}%
Using Eq.(\ref{GHF-f-s}) and Eq.(\ref{a1-b1-b2} ) we obtain the fundamental
set of solutions for the problem%
\begin{widetext}
\begin{gather}
w_{1}(t)=A_{_{-}1}F_{2}\left( \frac{1}{2}+\gamma ;\frac{1}{2}+\gamma
+i\delta ,\frac{1}{2}+\gamma -i\delta ;-\omega ^{2}e^{2t/T}\right)
+B_{-}(-\omega ^{2}e^{2t/T})^{\frac{1}{2}-\gamma -i\delta }\!_{1}F_{2}\left(
1-i\delta ;\frac{3}{2}-\gamma -i\delta ,1-2i\delta ;-\omega
^{2}e^{2t/T}\right) +  \notag \\
C_{-}(-\omega ^{2}e^{2t/T})^{\frac{1}{2}-\gamma +i\delta
}\!_{1}F_{2}\left( 1+i\delta ;1+2i\delta ,\frac{3}{2}-\gamma
+i\delta ;-\omega ^{2}e^{2t/T}\right) .  \notag \label{wtSol}
\end{gather}
\end{widetext}
In Eq.(\ref{wtSol}) $A_{-},$ $B_{-}$ and $C_{-}$ are integration constants.
Next step toward full solution is to determine the integration constants
from the initial conditions given by Eq.(\ref{u-v-w-initial}). Using Eq.(\ref%
{Bloch-eq}) it is straightforward to rewrite the initial conditions given by
Eq.(\ref{u-v-w-initial}) into%
\begin{equation}
w_{1}(-\infty )=-1,\ \dot{w}_{1}(-\infty )=0,\ \ddot{w}_{1}(-\infty )=0
\label{w-initial}
\end{equation}

From the transformation Eq.(\ref{x-t}) we observe that $x(-\infty )=0$ and
keep in mind Eq.(\ref{F(0)}) we will determine the integration constants $%
A_{-},$ $B_{-}$ and $C_{-}.$ One should note that the exponential factors $%
(-\omega ^{2}e^{2t/T})^{\frac{1}{2}-\gamma -i\delta }$ and $(-\omega
^{2}e^{2t/T})^{\frac{1}{2}-\gamma +i\delta }$ oscillate and when $1/2<\gamma
$ diverge in the limit $t\rightarrow -\infty $ $.$ In reason to have finite
value $w_{1}(-\infty )=-1$ we obtain%
\begin{equation}
A_{1}=-1,~B_{1}=C_{1}=0.  \label{A-B-C-initial}
\end{equation}

Finally in the interval $t\in I_{1}(-\infty ;0]$ the solution reads%
\begin{equation}
w_{1}(t)=-\!_{1}F_{2}\left( \frac{1}{2}+\gamma ;\frac{1}{2}+\gamma +i\delta ,%
\frac{1}{2}+\gamma -i\delta ;-\omega ^{2}e^{2t/T}\right) .
\label{w(t)-solution1}
\end{equation}%
Demkov model has a cusp for $\Omega (t)$ at $t=0$. This requires to derive a
solution for the interval $t\in I_{2}[0;+\infty )$, where the new initial
conditions at $t=0$ are obtained using Eq.(\ref{w(t)-solution1}) and Eq.(\ref%
{D-GHF}). We should stress that the initial condition given by Eq.(\ref{w1-2}%
) has not been derived by taking the second derivative of Eq.(\ref%
{w(t)-solution1}) at $t=0$. Because of the specific properties of the Demkov
model, i.e. cusp of the Rabi frequency $\Omega (t)$ at $t=0,$ one should
take the correct derivative of $\Omega (t)$ at $t\rightarrow 0_{+}$ and than
using the Bloch equations given by Eq.(\ref{Bloch-eq}), rigorously to obtain
the initial condition Eq.(\ref{w1-2}).

\begin{widetext}
\begin{eqnarray*}
w_{1}(0) &=&-\!_{1}F_{2}\left( \frac{1}{2}+\gamma ;\frac{1}{2}+\gamma
+i\delta ,\frac{1}{2}+\gamma -i\delta ;-\omega ^{2}\right) \newline
\\
\dot{w}_{1}(0) &=&\frac{2\left( \frac{1}{2}+\gamma \right) \omega ^{2}}{T%
\left[ \left( \frac{1}{2}+\gamma \right) ^{2}+\delta ^{2}\right] }%
\!_{1}F_{2}\left( \frac{3}{2}+\gamma ;\frac{3}{2}+\gamma +i\delta ,\frac{3}{2%
}+\gamma -i\delta ;-\omega ^{2}\right)  \\
\ddot{w}_{1}(0) &=&\frac{4\omega ^{2}\left( \frac{1}{2}+\gamma \right)
(1-T^{2})}{T^{2}\left[ \left( \frac{1}{2}+\gamma \right) ^{2}+\delta ^{2}%
\right] }\left[ _{1}F_{2}\left( \frac{3}{2}+\gamma ;\frac{3}{2}+\gamma
+i\delta ,\frac{3}{2}+\gamma -i\delta ;-\omega ^{2}\right) \right. + \\
&&\left. \frac{4\omega ^{2}\left( \frac{1}{2}+\gamma \right) \left( \frac{3}{%
2}+\gamma \right) }{T^{2}\left[ \left( \frac{1}{2}+\gamma \right)
^{2}+\delta ^{2}\right] \left[ \left( \frac{3}{2}+\gamma \right)
^{2}+\delta ^{2}\right] }\!_{1}F_{2}\left( \frac{5}{2}+\gamma
;\frac{5}{2}+\gamma +i\delta ,\frac{5}{2}+\gamma -i\delta ;-\omega
^{2}\right) \right] \label{w1-2}
\end{eqnarray*}
\end{widetext}
By analogy with Eq.(\ref{w-eq}) for the time interval $t\in I_{2}[0;+\infty
) $ we have the following equation for the population inversion
\begin{gather}
\dddot{w}_{2}+2(T^{-1}+\Gamma )\ddot{w}_{2}+\left[ (T^{-1}+\Gamma
)^{2}+\Delta ^{2}+\Omega _{0}^{2}e^{-2t/T}\right] \dot{w}_{2}+  \label{w2-eq}
\\
\Omega _{0}^{2}(-T^{-1}+\Gamma )e^{-2t/T}w_{2}=0.  \notag
\end{gather}%
The solution can again be expressed in term of GHF, after the transformation%
\begin{equation*}
x=-\frac{1}{4}(T\ \Omega _{0})^{2}e^{-2t/T}.
\end{equation*}%
The solution of the equation for the population inversion $w_{2}$ within the
time interval $t\in I_{2}[0;+\infty ),$ reads%
\begin{equation}
w_{2}(t)=A_{+}f_{1}+B_{+}f_{2}+C_{+}f_{3}.  \label{w2(t)-sol}
\end{equation}%
In this equation by reason to simplify some cumbersome formulas, we will
denote the three linearly independent solutions of Eq.(\ref{w2(t)-sol}) with
$f_{1},$ $f_{2}$ and $f_{3}.$ Adopting the notations introduced by Eq.(\ref%
{g-d-o}) $f_{1},$ $f_{2}$ and $f_{3}$ are given by
\begin{widetext}
\begin{subequations}
\label{f-all}
\begin{eqnarray}
f_{1}(t) &=&\!_{1}F_{2}\left( \frac{1}{2}-\gamma ;\frac{1}{2}-\gamma
-i\delta ,\frac{1}{2}-\gamma +i\delta ;-\omega ^{2}e^{-2t/T}\right)
\label{f1} \\
f_{2}(t) &=&\!(-\omega ^{2}e^{-2t/T})^{\frac{1}{2}+\gamma +i\delta
}\!_{1}F_{2}\left( 1+i\delta ;\frac{3}{2}+\gamma +i\delta ,1+2i\delta
;-\omega ^{2}e^{-2t/T}\right)  \label{f2} \\
f_{3}(t) &=&(-\omega ^{2}e^{-2t/T})^{\frac{1}{2}+\gamma -i\delta
}\!_{1}F_{2}\left( 1-i\delta ;1-2i\delta ,\frac{3}{2}+\gamma -i\delta
;-\omega ^{2}e^{-2t/T}\right)  \label{f3}
\end{eqnarray}%
\end{subequations}
\end{widetext}We have determined the integration constants $A_{-},$ $B_{-}$
and $C_{-}$ by using the asymptotic behaviours of the exponential factors $%
(-\omega ^{2}e^{2t/T})^{\frac{1}{2}-\gamma -i\delta }$ and $(-\omega
^{2}e^{2t/T})^{\frac{1}{2}-\gamma +i\delta }.$ This simple argumentation
cannot be used for Eq.(\ref{w2(t)-sol}) and respectively for $A_{+},$ $B_{+}$
and $C_{+}.$ We will determine the integration constants $A_{+},$ $B_{+}$
and $C_{+}$ using the initial conditions at $t=0$, given by Eq.(\ref%
{w1-2}). The textbook method requires to write a linear system of
equations for the unknown variables $A_{+},$ $B_{+}$ and $C_{+}$.
This is
done using Eq.(\ref{w2(t)-sol}) by taking $w_{2}(0)$ and the derivatives $%
\dot{w}_{2}(0)$ and $\ddot{w}_{2}(0).$ After straightforward albeit tedious
algebra, using Eq.(\ref{w1(0)-all}) and Eq.(\ref{f-all}) we obtain the
solution for the integration constants

\label{A2-B2-C2-all}
\begin{subequations}
\begin{eqnarray}
A_{+} &=&\frac{\left[
\begin{array}{ccc}
w_{1}(0) & f_{2}(0) & f_{3}(0) \\
\dot{w}_{1}(0) & \dot{f}_{2}(0) & \dot{f}_{3}(0) \\
\ddot{w}_{1}(0) & \ddot{f}_{2}(0) & \ddot{f}_{3}(0)%
\end{array}%
\right] }{W[f_{1}(0),f_{2}(0),f_{3}(0)]},  \label{A2} \\
B_{+} &=&\frac{\left[
\begin{array}{ccc}
f_{1}(0) & w_{1}(0) & f_{3}(0) \\
\dot{f}_{1}(0) & \dot{w}_{1}(0) & \dot{f}_{3}(0) \\
\ddot{f}_{1}(0) & \ddot{w}_{1}(0) & \ddot{f}_{3}(0)%
\end{array}%
\right] }{W[f_{1}(0),f_{2}(0),f_{3}(0)]},  \label{B2} \\
C_{+} &=&\frac{\left[
\begin{array}{ccc}
f_{1}(0) & f_{2}(0) & w_{1}(0) \\
\dot{f}_{1}(0) & \dot{f}_{2}(0) & \dot{w}_{1}(0) \\
\ddot{f}_{1}(0) & \ddot{f}_{2}(0) & \ddot{w}_{1}(0)%
\end{array}%
\right] }{W[f_{1}(0),f_{2}(0),f_{3}(0)]}.  \label{C2}
\end{eqnarray}%
The denominator of the expressions for $A_{+},$ $B_{+}$ and $C_{+}$ is the
Wronskian for the three linearly independent solutions $f_{1},$ $f_{2}$ and $%
f_{3}$.

Finally, we can write the solution for the Demkov model, bringing together
the results from Eqs.(\ref{w(t)-sol}), (\ref{A-B-C-initial}), (\ref%
{w2(t)-sol}), (\ref{f-all}) and (\ref{A2-B2-C2-all})
\end{subequations}
\begin{equation}
w(t)=w_{1}(t)\theta (-t)+w_{2}(t)\theta (t),  \label{w1+w2}
\end{equation}%
where $\theta (t)$ is the Heaviside "unit step" function. In the very some
manner the reader could derive solutions for the coherence components of the
Bloch vector $u(t)$ and $v(t).$ Hereafter our main concern will be to
investigate the solution for the population inversion $w(t).$

For variate of applications using exact soluble models, the final transition
probability expressed via $w(+\infty )$ is more important than time
dependent behaviour of the $w(t)$ itself. Having in mind the Demkov model
and its solution given by Eq.(\ref{w1+w2}), it is easy to be seen that only
the expression for $w_{2}(t)$ is important, when the final transition
probability is considered. A closed look to the three linearly independent
solutions $f_{1},$ $f_{2}$ and $f_{3}$ reveal that only $f_{1}$ term will
survive under the limit $t\rightarrow +\infty $. This is due to the
exponential factors $\!(-\omega ^{2}e^{-2t/T})^{\frac{1}{2}+\gamma +i\delta
} $ and $(-\omega ^{2}e^{-2t/T})^{\frac{1}{2}+\gamma -i\delta }$, which tend
to zero at $t\rightarrow +\infty $. Using Eq.(\ref{F(0)}) we arrive at the
following expression
\begin{equation}
w(+\infty )=A_{+}.  \label{w-infinity}
\end{equation}%
Although we have closed analytic result for the time dependent $w(t)$ as
well as $w(+\infty )$, their complexity impose the use of asymptotic
expressions for various limits.

\begin{figure}[tb]
\includegraphics[width=75mm]{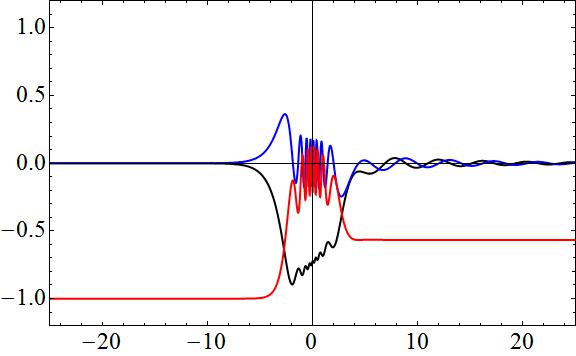}
\caption{Components of the Block vector. $u(t)$ is plotted with black curve;
$v(t)$ is plotted with blue curve; $w(t)$ is plotted with red curve; Model
parameters are the following $\protect\delta=1.5 $, $\protect\omega_{0}=25 $%
, $\protect\gamma=0.1 $}
\label{blockvector}
\end{figure}

Figure \ref{blockvector} displays the time evolution of the components of
the Block vector for a specific values of the parameters.

\section{Resonant coherent excitation}


Coherent resonant excitation represents an important notion in coherent
quantum processes. Besides from the requirement that the frequency of the
external field must be equal to the Bohr transition frequency i.e the
resonance condition, crucial condition for such processes is also the
coherence. Having derived the analytic solutions for the Demkov model in the
general case it is instructive to investigate the resonant regime. In this
simplified case the Bloch equations take the following form

\begin{eqnarray}
\frac{\text{d}}{\text{d}t}u^{r}(t) &=&-\Gamma u^{r}(t),
\label{Bloch-resonance-u} \\
\frac{\text{d}}{\text{d}t}\left[
\begin{array}{c}
v^{r}(t) \\
w^{r}(t)%
\end{array}%
\right] &=&\left[
\begin{array}{cc}
-\Gamma & -\Omega (t) \\
\Omega (t) & 0%
\end{array}%
\right] \left[
\begin{array}{c}
v^{r}(t) \\
w^{r}(t)%
\end{array}%
\right] .  \label{Bloch-resonance-u-w}
\end{eqnarray}%
In the last formula index "$r$" stands for resonant solution of the Bloch
vector. For the resonant case the Bloch equations factorize to a single
equation for the coherence $u^{r}(t)$ and a system of two equations for the
remaining components of the Bloch vector $\left[ v^{r}(t),w^{r}(t)\right]
^{T}$. Although the solution of the full Demkov problem requires three
linearly independent solutions written in Eqs.(\ref{f-all}), the solution
for the population inversion $w^{r}(t)$ in the resonant case, is derived
from Eq.(\ref{Bloch-resonance-u-w}). This system is reducible to a linear
differential equation of second order, that posses two linearly independent
solutions $f_{1}^{r}(t)$ and $f_{2}^{r}(t)$. Having in mind the symmetry of
the GHF and Eqs.(\ref{f-all}) it is seen that for the interval $t\in
I_{2}[0;+\infty )$, the solutions for the resonant case $\Delta =0,$ are
given by%
\begin{equation*}
f_{1}^{r}(t)=f_{1}(t)|_{\Delta
=0};~f_{2}^{r}(t)=f_{3}^{r}(t)=f_{2}(t)|_{\Delta =0}=f_{3}(t)|_{\Delta =0}
\end{equation*}
Furthermore the condition $\Delta =0,$ lead to significant simplification of
the GHF, which is given by the relation between the GHF and the Bessel
function

\begin{equation}
\!_{1}F_{2}\left( a;a,b;z\right) =\text{ }_{0}F_{1}\left( \text{ }%
;b;z\right) =\Gamma (b)\left( -z\right) ^{\frac{1-b}{2}}J_{b-1}(2\sqrt{-z}).
\label{GHF-Bessel}
\end{equation}%
Using Eq.(\ref{w(t)-sol}) and Eq.(\ref{GHF-Bessel}) we obtain the solution
for the resonant Demkov model within the interval $t\in I_{1}(-\infty ;0]$%
\begin{equation}
w_{1}^{r}(t)=-\ \Gamma \left( \frac{1}{2}+\gamma \right) \left( \omega \!\
e^{t/T}\right) ^{\frac{1}{2}+\gamma }J_{-\frac{1}{2}+\gamma }(2\omega \!\
e^{t/T}).  \label{wr1(t)-sol}
\end{equation}
By analogy with the initial conditions Eq.(\ref{w1(0)-all}) for the resonant
case we obtain \label{w1r(0)-all}
\begin{eqnarray}
w_{1}^{r}(0) &=&-\ \Gamma \left( \frac{1}{2}+\gamma \right) \omega \!^{\frac{%
1}{2}-\gamma }J_{-\frac{1}{2}+\gamma }(2\omega \!\ ),  \label{w1r-0} \\
\dot{w}_{1}^{r}(0) &=&\frac{2}{T\left( \frac{1}{2}+\gamma \right) }\!\Gamma
\left( \frac{3}{2}+\gamma \right) \omega \!^{\frac{3}{2}-\gamma }J_{\frac{1}{%
2}+\gamma }(2\omega \!\ ).  \label{w1r-1}
\end{eqnarray}

Having in mind the symmetry of the GHF, it is seen that for the interval $%
t\in I_{2}[0;+\infty )$, the solutions for the resonant case are given by
Eqs.(\ref{f-all}), where $f_{2}(t)=f_{3}(t),$ under the constrain $\Delta
=0. $ For the time interval $t\in I_{2}[0;+\infty )$ we have the following
solution of the equation for the population inversion
\begin{equation}
w_{2}^{r}(t)=A_{+}^{r}f_{1}^{r}+B_{+}^{r}f_{2}^{r}.  \label{wr2(t)-sol}
\end{equation}

In Eq.(\ref{wr2(t)-sol}) the following notation were used \label{fr-all}
\begin{eqnarray}
f_{1}^{r}(t) &=&\ \Gamma \left( \frac{1}{2}-\gamma \right) \left( \omega \!\
e^{-t/T}\right) ^{\frac{1}{2}+\gamma }J_{-\frac{1}{2}-\gamma }(2\omega \!\
e^{-t/T}),  \label{f1r} \\
f_{2}^{r}(t) &=&\!(-1)^{\frac{1}{2}+\gamma }\!\left( \omega \!\
e^{-t/T}\right) ^{\frac{3}{2}+3\gamma }\Gamma \left( \frac{1}{2}-\gamma
\right) J_{-\frac{1}{2}-\gamma }(2\omega \!\ e^{-t/T}).  \label{f2r}
\end{eqnarray}%
Integration constants $A_{+}^{r}$ and $B_{+}^{r}$ are given by \label%
{A2r-B2r}
\begin{eqnarray}
A_{+}^{r} &=&\frac{\left[
\begin{tabular}{ll}
$w_{1}^{r}(0)$ & $f_{2}^{r}(0)$ \\
$\dot{w}_{1}^{r}(0)$ & $\dot{f}_{2}^{r}(0)$%
\end{tabular}%
\right] }{W[f_{1}^{r}(0),f_{2}^{r}(0)]},  \label{A2r} \\
B_{+}^{r} &=&\frac{\left[
\begin{tabular}{ll}
$f_{1}^{r}(0)$ & $w_{1}^{r}(0)$ \\
$\dot{f}_{1}^{r}(0)$ & $\dot{w}_{1}^{r}(0)$%
\end{tabular}%
\right] }{W[f_{1}^{r}(0),f_{2}^{r}(0)]}.  \label{B2r}
\end{eqnarray}

\section{Conclusions}


\acknowledgments This work has been supported by the project QUANTNET -
European Reintegration Grant (ERG) - PERG07-GA-2010-268432.



\section{APPENDIX}


For the sake of readers convenience we summarize here some relevant
properties of the GHF. Further details can be found in \cite{Rainville}.
Generalized hypergeometric function $_{1}F_{2}(a_{1};b_{1},b_{2};x)$ can be
introdused as a power series
\begin{equation}
_{1}F_{2}(a_{1};b_{1},b_{2};x)=\sum_{k=0}^{\infty }\frac{(a_{1})_{k}\ }{%
(b_{1})_{k}(b_{2})_{k}}\frac{x^{k}}{k!}.  \label{GHF-series}
\end{equation}%
Here it is assumed that none of the bottom parameters $b_{1}$ and $b_{2}$ is
a nonpositive integer. As usual $(a_{1})_{k}$ , $(b_{1})_{k}$ and $%
(b_{2})_{k}$ are Pochhamer symbols i.e. $(\alpha )_{0}=1$ and $(\alpha
)_{k}=\alpha (\alpha +1)...(\alpha +k-1)$. Series given by Eq.(\ref%
{GHF-series}) converges for all finite values of $x$ and defines an entire
function. Generalized hypergeometric function $%
_{1}F_{2}(a_{1};b_{1},b_{2};x) $ satisfies the differential equation Eq.(\ref%
{GHF-eq}). When neither $b_{1}$ and $b_{2\text{ }}$are integers, nor the
difference $b_{1}-b_{2}$, a fundamental set of solutions of Eq.(\ref{GHF-eq}%
) is given by%
\begin{eqnarray}
w(x) &=&A_{1}F_{2}(a_{1};b_{1},b_{2};x)+  \label{GHF-f-s} \\
&&B\ x^{1-b_{1}}{}_{1}F_{2}(a_{1}+1-b_{1};2-b_{1},b_{2}+1-b_{1};x)+ \\
&&C\ x^{1-b_{2}}{}_{1}F_{2}(a_{1}+1-b_{2};b_{1}+1-b_{2},2-b_{2};x)  \notag
\end{eqnarray}%
We have, in the neighborhood of the origin three linearly independent
solutions. It can be shown \cite{Rainville} that the Wronskian of these
solutions is given by%
\begin{widetext}
\begin{eqnarray*}
&&W\left[
_{1}F_{2}(a_{1};b_{1},b_{2};x);x^{1-b_{1}}{}_{1}F_{2}(a_{1}+1-b_{1};2-b_{1},b_{2}+1-b_{1};x);x^{1-b_{2}}{}_{1}F_{2}(a_{1}+1-b_{2};b_{1}+1-b_{2},2-b_{2};x)%
\right]  \\
&=&(b_{1}-1)(b_{2}-1)(b_{1}-b_{2})x^{-b_{1}-b_{2}-1}
\end{eqnarray*}
\end{widetext}
Some usefull formulas could be derived from the deffinition Eq.(\ref%
{GHF-series}). It follows that%
\begin{equation}
_{1}F_{2}(a_{1};b_{1},b_{2};0)=1  \label{F(0)}
\end{equation}%
The derivatives for the GHF with respect of the independent variable $x$ are
given by
\begin{equation}
\frac{d\!\!^{n}}{dx^{n}}\!_{1}F_{2}(a_{1};b_{1},b_{2};x)=\frac{(a_{1})_{n}\
}{(b_{1})_{n}(b_{2})_{n}}\!_{1}F_{2}(a_{1}+n;b_{1}+n,b_{2}+n;x)
\label{D-GHF}
\end{equation}%
Asymtotic expansion of the GHF is given by%
\begin{gather*}
_{1}F_{2}(a_{1};b_{1},b_{2};z)\symbol{126}\frac{\Gamma (b_{1})\Gamma
(b_{2})\ }{2\sqrt{\pi }\Gamma (a_{1})}(-z)^{\chi }\cos (\pi \chi +2\sqrt{-z}%
)\left( 1+O(1/\sqrt{-z})\right) + \\
\frac{\Gamma (b_{1})\Gamma (b_{2})\ }{\Gamma (b_{1}-a_{1})\Gamma
(b_{2}-a_{1})}(-z)^{-a_{1}}\left( 1+O(1/z)\right) ,\text{ \ }|z|\rightarrow
\infty
\end{gather*}%
where
\begin{equation*}
\chi =\frac{1}{2}\left( a_{1}-b_{1}-b_{2}+\frac{1}{2}\right)
\end{equation*}

\end{document}